\documentclass[aps,prl,twocolumn,footinbib,superscriptaddress]{revtex4-2}
\usepackage{graphicx}
\usepackage{amsmath}
\usepackage{mathtools}
\usepackage{amssymb}

\DeclareMathOperator{\re}{Re}
\DeclareMathOperator{\im}{Im}
\newcommand{\vmin}{v_\text{min}}
\newcommand{\vmax}{v_\text{max}}
\newcommand{\I}{\text{I}}
\newcommand{\II}{\text{II}}
\newcommand{\R}{\mathbb{R}}
\newcommand{\C}{\mathbb{C}}

\begin{document}

\title{Two-phase quadratic integrate-and-fire neurons:\\ Exact low-dimensional description for ensembles of finite-voltage neurons}
\author{Rok Cestnik}
\affiliation{Centre for Mathematical Sciences, Lund University, M\"arkesbacken 4, 22362 Lund, Sweden}
\email{rok.cestnik@math.lth.se}

\begin{abstract}
We introduce a two-phase quadratic integrate-and-fire (QIF) neuron whose membrane potential evolves according to two alternating Riccati equations within finite bounds.
This simple extension removes the unphysical voltage divergence of the standard QIF model while producing realistic spike waveforms. 
Despite this modification, the system retains an exact low-dimensional description in the thermodynamic limit, governed by a single complex Riccati equation. 
Expressions for collective quantities such as the firing rate and mean voltage remain compact and analytically tractable. 
Because the formalism preserves the mathematical structure of the standard QIF ensemble, it inherits its many generalizations and can serve as a drop-in replacement in existing mean-field frameworks, providing a more biologically plausible yet still exactly solvable neuronal model.
\end{abstract}

\maketitle

The quadratic integrate-and-fire (QIF) neuron~\cite{ermentrout_kopell_1986,ermentrout_1996,latham_et_al_2000} is a minimal spiking model whose population activity admits an exact low-dimensional mean-field description via the Lorentzian / Ott-Antonsen ansatz~\cite{montbrio_pazo_roxin_2015,ott_antonsen_2008}.
This analytical tractability has made QIF ensembles~\cite{pietras_montbrio_2024,bi_et_al_torcini_2021,clusella_montbrio_2024,divolo_torcini_2018,pietras_et_al_montbrio_2019,montbrio_pazo_2020,ratas_pyragas_2016,coombes_2023,goldobin_divolo_torcini_2021,goldobin_2021,goldobin_et_al_klimenko_2024,taher_torcini_olmi_2020,coombes_byrne_2019,byrne_et_al_coombes_2020} and their equivalent $\theta$-neuron formulations~\cite{luke_et_al_so_2013,so_et_al_barreto_2014,laing_2014,laing_2015}, as well as the more general phase-oscillator ensembles~\cite{pazo_montbrio_2014,tyulkina_et_al_pikovsky_2018,cestnik_pikovsky_2022,cestnik_pikovsky_2022b,pikovsky_rosenblum_2015,skardal_2018,pyragas_pyragas_2022,pikovsky_rosenblum_2011,seth_et_al_strogatz_2009,martens_et_al_antonsen_2009,omelchenko_2023,laing_2009,omelchenko_et_al_sudakov_2012}, 
the standard tools for studying collective dynamics. 
Exact low-dimensional descriptions are rare and particularly valuable~\cite{ashwin_et_al_nicks_2016,bick_et_al_martens_2020}, to the extent that many works seek to reproduce their structure only approximately in more realistic neuron models with adaptation or synaptic plasticity, by introducing additional mean-field or mesoscopic assumptions that retain analytical tractability~\cite{gast_et_al_knosche_2020,gast_et_al_schmidt_2021,ferrara_et_al_olmi_2023,gast_et_al_kennedy_2023,chen_campbell_2022,duchet_bick_byrne_2023,fennelly_et_al_byrne_2025}.

A well-known limitation of the QIF model is that its voltage variable diverges at spike emission, which complicates its biological interpretation.
We introduce a two-phase generalization of the QIF neuron that removes this divergence while preserving exact integrability. 
In this model, a second phase bounds the membrane potential but still permits a closed-form mean-field reduction in the thermodynamic limit. 
As a result, we obtain population equations of QIF type, with richer and more biologically plausible voltage dynamics.

\emph{Two-phase quadratic integrate-and-fire neuron. }
The voltage $v_j$ evolves within a bounded interval $v_j \in [\vmin, \vmax]$, where $\vmin < 0$ and $\vmax > 0$. 
Each neuron alternates between two phases, denoted $v_j^\I$ and $v_j^\II$. 
In both phases, the voltage follows a real-valued Riccati equation with time-varying global coefficients $(a,b,c)$:
\begin{equation}\label{eq:micro}
\dot{v}_j = a^{(p)} v_j^2 + b^{(p)} v_j + c^{(p)}\,,
\end{equation}
where $p \in \{\I, \II\}$ denotes the phase and index $j = 1, \dots, N$. 
A neuron switches phase whenever its voltage reaches one of the boundaries, 
i.e. when $v_j = \vmax$ or $v_j = \vmin$.

\begin{figure}[ht]
	\centering
	\includegraphics[width=0.99\columnwidth]{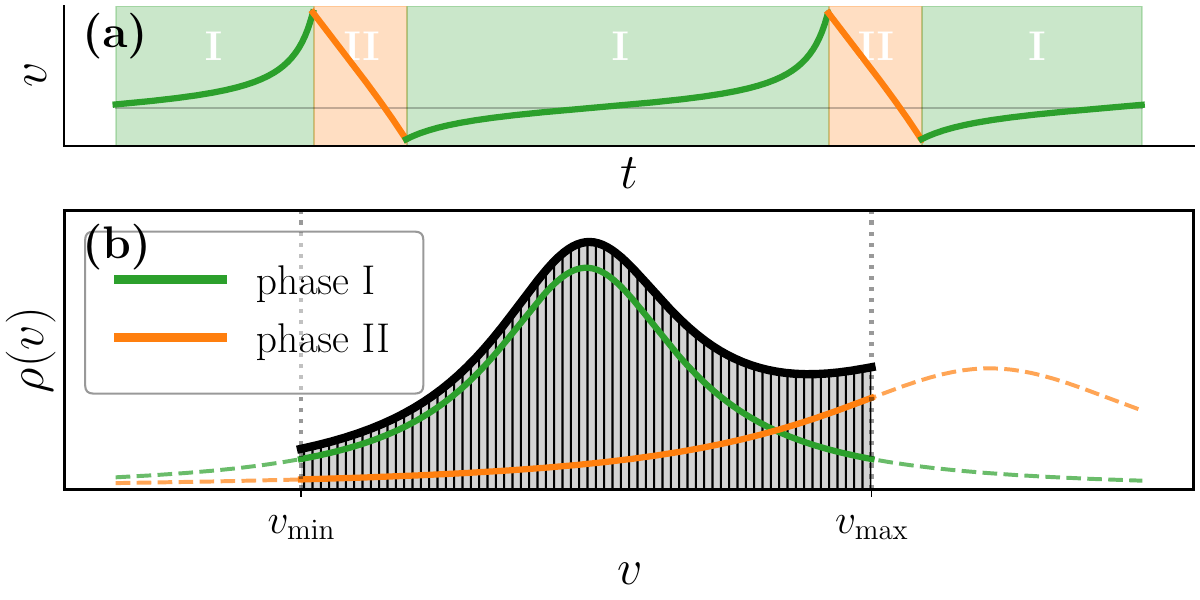}
	\caption{Schematic representation of a single two-phase QIF~\eqref{eq:micro} voltage time series in panel $(a)$ and the ansatz distribution~\eqref{eq:distribution} in panel $(b)$. Contributions of the first phase are marked with green and of the second phase with orange. In panel $(b)$ the full distribution $\rho(v)$ is depicted with black. }
	\label{fig1}
\end{figure}

\begin{figure}[ht]
	\centering
	\includegraphics[width=0.99\columnwidth]{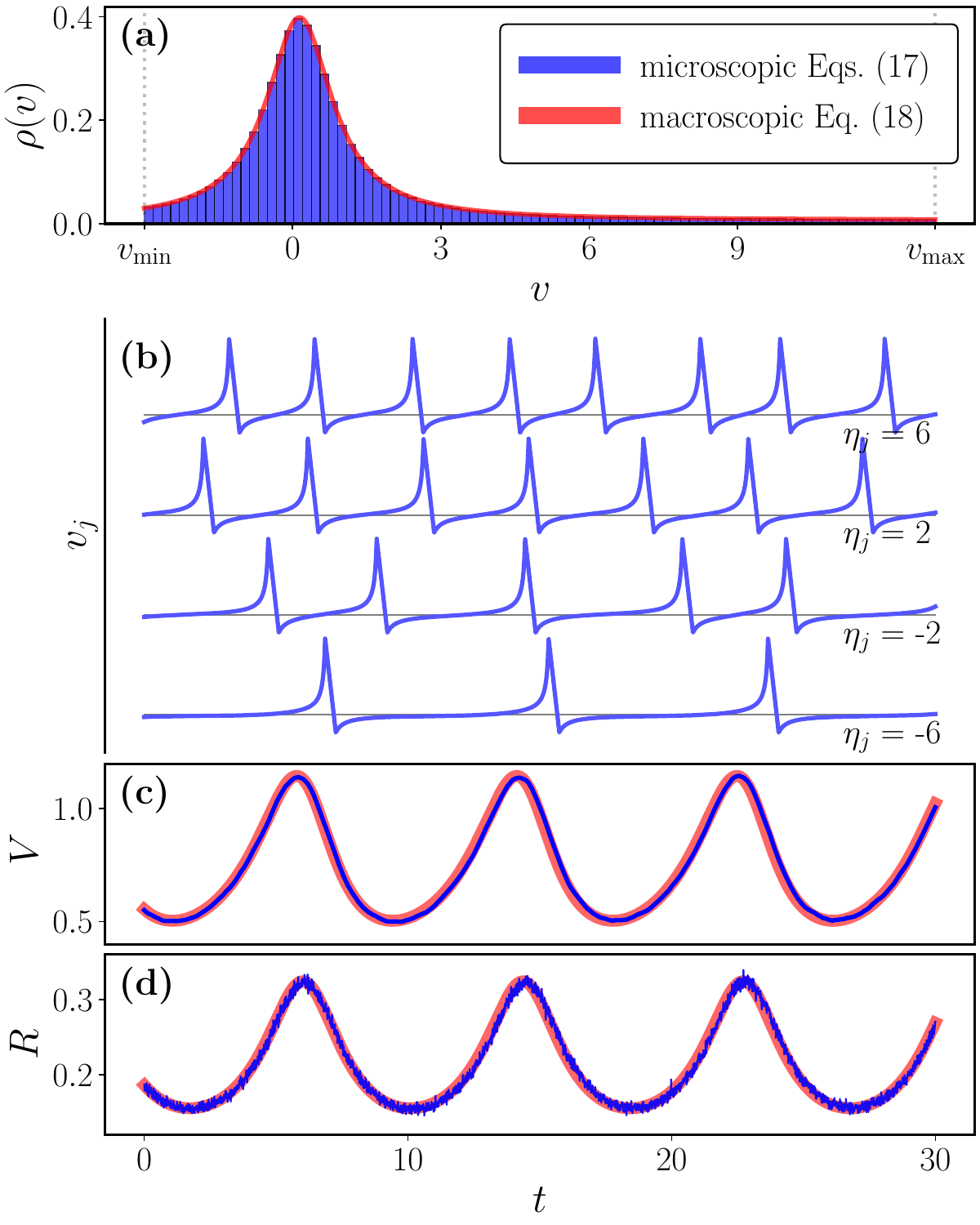}
	\caption{Example simulation comparing the dynamics from a large ($N=10^6$), finite ensemble of microscopic equations~\eqref{eq:micro_ex} (blue) and exact macroscopic equation~\eqref{eq:macro_ex} (red). In panel $(a)$ the distribution of voltages~\eqref{eq:distribution}, in panel $(b)$ voltage traces for some values of the heterogeneity $\eta_j$, and in panels $(c)$ and $(d)$ the mean voltage $V$ and firing rate $R$ respectively in time. Parameters are: $\vmin=-3$, $\vmax=13$, $J=3$, $g=0.05$, $I=-0.2$, $\eta_0=0$, and $\Delta=0.05$. }
	\label{fig2}
\end{figure}

The coefficients of one phase can be chosen arbitrarily; 
without loss of generality, we take the first phase to have prescribed coefficients 
($a^{\I}, b^{\I}, c^{\I}$). 
The coefficients of the second phase ($a^{\II}, b^{\II}, c^{\II}$) 
are then fully determined by coefficients of the first phase and voltage boundaries $\vmin$, $\vmax$:
\begin{equation}\label{eq:abc}
\begin{aligned}
a^\II &= \frac{c^\I}{\vmin\vmax}\,,\\
b^\II &= -b^\I - c^\I \frac{2(\vmin+\vmax)}{\vmin\vmax}\,,\\
c^\II &= a^\I \vmin\vmax + b^\I (\vmin+\vmax) + c^\I \frac{(\vmin+\vmax)^2}{\vmin\vmax}\,.
\end{aligned}
\end{equation}
This construction ensures the voltage remains continuous at the junctions between phases and, 
for suitable choices of $\vmin$ and $\vmax$, produces realistic spike profiles 
[see Figs.~\ref{fig1}$(a)$~and~\ref{fig2}$(b)$ for examples].

\emph{Ansatz. }
We consider an infinite ensemble of neurons, $N \to \infty$, 
whose state is described by the probability density function of voltages, $\rho(v)$. 
The Lorentzian ansatz~\cite{montbrio_pazo_roxin_2015} generalizes to the two-phase QIF model 
as the sum of two truncated Lorentzian contributions, one for each phase:
\begin{equation}\label{eq:distribution}
\begin{aligned}
\rho(v) &= \rho^\I(v) + \rho^\II(v)\,, \qquad \text{for}\ v\in[\vmin,\vmax]\,, \\
\rho^\I(v) &= \frac{1}{\pi} \frac{\im[Q^\I]}{(v-\re[Q^\I])^2+\im[Q^\I]^2}\,, \\
\rho^\II(v) &= \frac{1}{\pi} \frac{\im[Q^\II]}{(v-\re[Q^\II])^2+\im[Q^\II]^2}\,,
\end{aligned}
\end{equation}
where $Q^\I \in \C$ is a complex-valued macroscopic quantity that parametrizes the distribution, while $Q^\II$ is determined from $Q^\I$ via:
\begin{equation}\label{eq:QII_def}
Q^\II = -\frac{\vmin \vmax}{\bar Q^\I}+\vmin+\vmax\,,
\end{equation}
similar to how the forces in the second phase ($a^\II,b^\II,c^\II$) are determined by those of the first ($a^\I, b^\I, c^\I$). 
The imaginary part of $Q^\I$ should be positive by convention, since it represents the width of the truncated Lorentzian distribution. 
Notice that $Q^\I$ is complex conjugated in expression~\eqref{eq:QII_def}, to ensure that the imaginary part of $Q^\II$ is also positive (representing the width of the second truncated Lorentzian).  

Note that the ansatz distribution~\eqref{eq:distribution} requires no additional normalization factors, 
even though each Lorentzian contribution is defined only over the finite interval $[\vmin,\vmax]$ --- 
this is not a coincidence, the two phases perfectly complement each other.
This type of distribution is illustrated in Figs.~\ref{fig1}$(b)$~and~\ref{fig2}$(a)$~\footnote{This type of distribution~\eqref{eq:distribution} can in practice be sampled by drawing standard Cauchy random numbers corresponding to parameter $Q$: $C \sim \mathrm{Cauchy}(\re[Q], \im[Q])$ 
(for example, by taking the ratio of two Gaussian random numbers) and if the sampled value falls outside the interval $[\vmin,\vmax]$, it is transformed via Eq.~\eqref{eq:transform}: $-\vmin\vmax/C+\vmin+\vmax$.}.

\emph{Exact macroscopic equations. }
The exact macroscopic dynamics follow a complex Riccati equation:
\begin{equation}\label{eq:macro}
\dot{Q} = aQ^2 + bQ + c\,,
\end{equation}
where $Q\coloneqq Q^\I$ is the complex quantity parametrizing the ansatz distribution~\eqref{eq:distribution} and coefficients $(a,b,c) \coloneqq (a^\I,b^\I,c^\I)$ are those of the first phase. 
Note that these dynamics have the same form as those of the standard Lorentzian ansatz~\cite{montbrio_pazo_roxin_2015}, but the quantity $Q\in\C$ now has a different meaning, and coefficients $(a,b,c)$ contain different coupling terms. 
In Ref.~\cite{montbrio_pazo_roxin_2015}, the real and imaginary parts of $Q$ would correspond to mean voltage $V$ and firing rate $R$ ($V = \re[Q], R = a\im[Q]/\pi$), but in the two-phase QIF ensemble these quantities are harder to express, yet still depend only on $Q$ [we express them later on, see Eqs.~\eqref{eq:R_def}~and~\eqref{eq:V_def}]. 

For a population of identical neurons, the coefficients $(a,b,c)$ entering Eq.~\eqref{eq:macro} are real and exactly the same as those of the first phase voltage equation~\eqref{eq:micro}. 
We will later show however, how heterogeneous ensembles can be described within the same framework as well, 
and the heterogeneity results in the addition of imaginary components to coefficients in the macroscopic dynamics equation~\eqref{eq:macro}.

\emph{Why does this work? }
The reason this construction works is that the system is equivalent to having a single voltage variable 
that is transformed whenever it leaves the interval $[\vmin, \vmax]$. 
Imagine two voltage variables that follow exclusively one of the dynamical phases; let us denote them as $v^\I$ and $v^\II$. Either of them can be considered the main variable, and they are just transformations of each other.  

If the main variable is the phase~$\I$ voltage $v^\I$, then outside the range $[\vmin,\vmax]$ voltage $v^\II$ can be expressed with the transformation: 
\begin{equation}\label{eq:transform}
v_j^\II = -\frac{\vmin \vmax}{v_j^\I}+\vmin+\vmax\,, \qquad (\text{for}\ v_j^\I\notin[\vmin,\vmax])\,.
\end{equation}
If the roles are reversed, the transformation is:
\begin{equation}\label{eq:transform_inv}
v_j^\I = \frac{\vmin\vmax}{\vmin+\vmax-v_j^\II}\,, \qquad (\text{for}\ v_j^\II\notin[\vmin,\vmax])\,.
\end{equation}
This transformation was chosen such that voltage is continuous along the seam of the two phases. 
In principle, one could use any M\"obius transform but then the voltage would be discontinuous at the junction.

\emph{How does this change the model from standard QIF ensembles? }
First, note that for uncoupled neurons, the model preserves the same intrinsic oscillatory structure as the standard QIF neuron: the period is identical and the first phase coincides with the QIF voltage dynamics. The inclusion of a second phase modifies the spike waveform, yielding finite voltages throughout the cycle, but does not change the fundamental timing of the oscillation.
The main differences therefore arise in the coupling, which we formulate in terms of two macroscopic quantities:
\begin{enumerate}
\item Firing rate $R$: the firing events now occur at a finite voltage $\vmax < \infty$~\footnote{Defining spikes as voltage threshold crossings can, in rare cases, lead to counting the same spike multiple times if the voltage during phase~$\II$ exceeds $\vmax$, effectively looping back through the firing threshold. This artifact is only relevant for small threshold values of $\vmax$.}.
\item The mean voltage $V$: now expresses differently due to the two-phase structure of the distribution $\rho(v)$~\eqref{eq:distribution}.
\end{enumerate}

Regarding point 1: the firing rate is now given by the flux at the finite voltage $\vmax$:
\begin{equation}
R = \left(\dot{v} \rho(v) \right)\Big|_{\vmax} = \frac{1}{\pi}\frac{\im[Q] (a \vmax^2 + b \vmax + c)}{(\vmax-\re[Q])^2+\im[Q]^2}\,.
\end{equation}
This can be rewritten without coefficients $b$ and $c$ in terms of the temporal derivative $\dot{Q}$:
\begin{equation}\label{eq:R_direct}
\pi R = a\im[Q]+\frac{\im[(\vmax-\bar Q)\dot{Q}]}{|\vmax-Q|^2}\,.
\end{equation}
Since the firing rate $R$ will enter equation~\eqref{eq:macro}, the dependence on $\dot{Q}$ 
might suggest that the resulting system is implicit, but this is not the case. 
We will here only consider firing rate coupling where $R$ enters additively in the voltage equation, i.e. is added to the $c^\I$ coefficient as: 
\begin{equation}\label{eq:f_def}
\dot{Q} = f(Q) + J R\,,
\end{equation}
where $J$ denotes the (chemical) coupling strength, and all remaining terms in Eq.~\eqref{eq:macro} (that do not depend on $R$) have been collected into $f(Q)$. 
Thus, we can solve for $R$ explicitly, by combining Eqs.~\eqref{eq:R_direct} and~\eqref{eq:f_def}:
\begin{equation}\label{eq:R_def}
R = \frac{ a\im[Q]|\vmax-Q|^2 + \im[(\vmax-\bar Q)f(Q)] }{ \pi|\vmax-Q|^2 - J \im[Q] }\,.
\end{equation}
Note that for stationary states, where $\dot{Q} = 0$, the firing rate coincides 
with that of the standard single-phase QIF ensemble: $R = a\im[Q] / \pi$. 
However, the parameter regions corresponding to stationary and non-stationary states differ between the standard QIF model and its two-phase extension. 

Regarding point 2: 
the mean of a truncated Lorentzian with center $\re[Q]$ and half width at half maximum (HWHM) $\im[Q]$ is:
\begin{equation}\label{eq:E_minor}
E(v | \vmin<v<\vmax) = \frac{\im\left[ Q \log\left( \frac{Q-\vmax}{Q-\vmin} \right) \right]}{\arg\left( \frac{Q-\vmax}{Q-\vmin} \right)}\,,
\end{equation}
where $\arg(\cdot)$ is the complex argument function. 
Additionally we have to determine the proportion of neurons within each phase, which is the proportion of a Lorentzian between the same boundaries:
\begin{equation}\label{eq:P_minor}
P(v | \vmin<v<\vmax) = \frac{1}{\pi} \arg\left( \frac{Q-\vmax}{Q-\vmin} \right)\,. 
\end{equation}
The mean voltage $V$ is determined by the product of Eqs.~\eqref{eq:E_minor} and~\eqref{eq:P_minor} for contributions of both phases, conveniently canceling the $\arg(\cdot)$ function:
\begin{equation}\label{eq:V_def}
V = \frac{1}{\pi}\im\left[ Q \log\left( \frac{Q-\vmax}{Q-\vmin} \right) + Q^\II \log\left( \frac{Q^\II-\vmax}{Q^\II-\vmin} \right) \right]\,,
\end{equation}
where $Q^\II$ is defined in Eq.~\eqref{eq:QII_def}. 
Higher moments of the distribution can also be expressed analytically~\footnote{
\mbox{Second moment: 
$V_2 = \langle v_j^2 \rangle = \frac{1}{\pi} \im\left[F_2(Q) + F_2(Q^\II)\right]$, } 
where $F_2(Q) = Q(\vmax-\vmin)+Q^2\log\left(\frac{Q-\vmax}{Q-\vmin}\right)$.
\mbox{Third moment: 
$V_3 = \langle v_j^3 \rangle = \frac{1}{\pi}\im\left[ F_2(Q) + F_3(Q^\II) \right]$, }
where $F_3(Q) = Q\frac{(\vmax^2-\vmin^2)}{2}+Q^2(\vmax-\vmin)+Q^3\log\left(\frac{Q-\vmax}{Q-\vmin}\right)$}.

\emph{Addition of heterogeneity. }
Consider now that the coefficients $(a,b,c)$ of both phases can depend on the neuron index $j$. 
We focus on the special case of the Lorentzian-distributed heterogeneity, because this case is elegantly admissible to exact formulation.  
Adding Lorentzian quenched heterogeneity $\eta_j\in\R$, with mean $\eta_0\in\R$ and HWHM $\Delta\in\R$, leads to the appearance of the complex term $\eta_0+i\Delta$ on the right-hand side of the $Q$ equation~\eqref{eq:macro}, where $i\coloneqq\sqrt{-1}$ is the imaginary unit. 
For instance, adding $\eta_j$ to $c^\I$ in Eq.~\eqref{eq:micro} results in $\eta_0+i\Delta$ being added to $c$ in Eq.~\eqref{eq:macro}. 

This is fully analogous to the contour-integral argument in the Ott-Antonsen ansatz~\cite{ott_antonsen_2008} and the Lorentzian ansatz~\cite{montbrio_pazo_roxin_2015}:
for each value of the heterogeneity $\eta$ consider there is a subpopulation of neurons that falls into the ansatz distribution~\eqref{eq:distribution} parametrized by the local order parameter $q(\eta)$. 
The global order parameter $Q$ is then obtained as the integral of all $q(\eta)$ over the distribution of heterogeneities $h(\eta)$: 
\begin{equation}
Q = \int\limits_{-\infty}^\infty q(\eta) h(\eta) d\eta\,.
\end{equation}
If $h(\eta)$ is Lorentzian with mean $\eta_0$ and HWHM $\Delta$, and $q(\eta)$ is analytic in the upper $\eta$ complex half-plane, this integral can be evaluated via the residue theorem. It is equal to the value of $q(\eta)$ at the Lorentzian pole $\eta = \eta_0+i\Delta$:
\begin{equation}
Q = \oint q(\eta) h(\eta) d\eta = q(\eta_0+i\Delta)\,.
\end{equation}
For this evaluation to be valid, $q(\eta) h(\eta)$ must vanish as $|\eta|\to\infty$, ensuring the contribution from the contour arc is zero. 
For Riccati equations like~\eqref{eq:macro}, one can verify that $|q(\eta)| \sim \sqrt{|\eta|}$ for $|\eta|\to\infty$, which satisfies this requirement. 

Important note --- this only works when heterogeneity $\eta_j$ is considered to act differently on both phases, i.e. according to transformations~\eqref{eq:abc}. Otherwise, if $\eta_j$ is added to both $c^\I$ and $c^\II$ in the same way, this description~\eqref{eq:macro} is no longer exact. 

\emph{Example. }
Following standard QIF network formulations, we incorporate chemical synaptic coupling of strength $J$ through the population firing rate and electrical coupling of strength $g$ through gap-junction-mediated voltage diffusion~\cite{laing_2015}. The voltage $v_j$ in the first phase then evolves according to:
\begin{equation}\label{eq:micro_ex}
\dot{v}_j = v_j^2 + I + J R + g(V-v_j) + \eta_j\,,
\end{equation}
where $R$ is the firing rate~\eqref{eq:R_def}, $V$ is the mean voltage~\eqref{eq:V_def}, and $\eta_j$ the Lorentzian heterogeneity. 
For each neuron $j$, the coefficients of the second phase ($a^\II_j$, $b^\II_j$, $c^\II_j$) are then determined by the transformation~\eqref{eq:abc} of coefficients of the first phase: $a^\I = 1$, $b^\I = -g$, $c_j^\I = I + J R + g V + \eta_j$. 
Note how in the first phase only $c^\I_j$ coefficient is heterogeneous, but through the transformation~\eqref{eq:abc} all the coefficients of the second phase depend on the neuron index $j$. 

In the low-dimensional description, the macroscopic equation has the same functional form as the microscopic phase~$\I$ voltage equation~\eqref{eq:micro_ex}, with the microscopic variable $v_j$ replaced by the macroscopic variable $Q\in\C$ and the heterogeneities $\eta_j\in\R$ entering as the complex term $\eta_0 + i\Delta$, where $\eta_0\in\R$ and $\Delta\in\R$ denote the center and HWHM of the Lorentzian heterogeneity distribution:
\begin{equation}\label{eq:macro_ex}
\dot{Q} = Q^2 + I + J R + g(V-Q) + \eta_0 + i\Delta\,.
\end{equation}

Figure~\ref{fig2} shows simulation results comparing a large ensemble of $N=10^6$ microscopic voltage equations~\eqref{eq:micro_ex} to the exact macroscopic equation~\eqref{eq:macro_ex}. 
The two descriptions match exactly, as expected from the theory.
The parameters used are $\vmin=-3$, $\vmax=13$, $I=-0.2$, $J=3$, $g=0.05$, $\eta_0=0$, and $\Delta=0.05$, for which the ensemble exhibits periodic collective motion --- for the same parameter values $(I,J,g,\eta_0)$, the standard QIF ensemble exhibits only asymptotically stationary states. 
The simulation code is available in Ref.~\cite{fig2_code}. 

\emph{Discussion. }
We introduced the two-phase QIF neuron, whose membrane potential evolves over two alternating phases and is bounded between 
$\vmin$ and $\vmax$. 
This construction yields realistic spike shapes while retaining the exact mean-field reduction of standard QIF ensembles. 
Because the two phases follow different dynamics, coupling influences the system differently across them and is effectively weakened during the spike phase. This weakening can be understood from the construction of the model: the recovery phase is obtained as a smooth transformation~\eqref{eq:transform} of the standard QIF voltage near the spike, where the intrinsic dynamics dominate and external inputs have little influence on the intrinsic timing of the trajectory. 
As a consequence, the reduced sensitivity to inputs --- characteristic of the standard QIF near spikes --- carries over to the two-phase model, despite the voltage remaining bounded. This reflects a realistic feature of neuronal behavior --- an effective insensitivity to inputs during spikes --- that emerges naturally in this framework. 

Since the voltage now follows realistic spike trajectories, explicit pulse coupling --- implemented as discrete, instantaneous interactions at spike times --- may appear less natural in this setting. Instead, the mean voltage $V$ can serve as a continuous proxy for synaptic activity, closely tracking the firing rate $R$ up to a scaling factor [compare Fig.~\ref{fig2} panels~$(c)$~and~$(d)$]. 

Previous works have addressed related limitations of the standard QIF model.
Reference~\cite{montbrio_pazo_2020} introduced asymmetric reset rules and derived an exact reduction valid only in the limit of diverging voltages; for finite resets, their voltage profiles resemble those of the present two-phase model (cf. their Fig.~1), though the reduction is no longer exact.
Reference~\cite{pietras_2024} developed a framework that allows exact reductions for ensembles with prescribed pulse shapes. 
In contrast, in the two-phase QIF model, the realistic spike waveform emerges naturally from the microscopic voltage equations rather than being imposed externally.

The two-phase QIF model also inherits the extensive generalizations developed for standard QIF ensembles.
The formalism extends beyond the Lorentzian-like ansatz~\eqref{eq:distribution}, allowing any initial voltage distribution to be evolved exactly~\cite{pietras_cestnik_pikovsky_2023, cestnik_pikovsky_2022, cestnik_pikovsky_2022b}.
Different heterogeneity distributions can likewise be incorporated using the general frameworks proposed in Refs.~\cite{skardal_2018, pyragas_pyragas_2022}, which provide systematic ways to approximate various distribution forms. 
Moreover, the clever inclusion of adaptation mechanisms in Ref.~\cite{pietras_et_al_montbrio_2025} is directly compatible with the present framework.
In this sense, the two-phase QIF model can serve as a drop-in replacement for the standard QIF neuron in established low-dimensional frameworks, while providing bounded voltages and realistic spike waveforms.
Just as in standard QIF ensembles, the ansatz distribution~\eqref{eq:distribution} is invariant under the dynamics, but even arbitrarily weak heterogeneity renders it attracting~\cite{pietras_cestnik_pikovsky_2023,pietras_daffertshofer_2016,engelbrecht_mirollo_2020}, so that any initial voltage distribution asymptotically converges to it.

Cauchy noise added to the standard QIF ensemble has the same macroscopic effect as Lorentzian heterogeneity~\cite{toenjes_pikovsky_2020, tanaka_2020, pietras_cestnik_pikovsky_2023, pyragas_pyragas_2023, clusella_montbrio_2024}. 
In the two-phase QIF framework however, this correspondence breaks: it would be exact if the noise entered according to the phase transformation~\eqref{eq:abc} --- as is the case for heterogeneity --- but for independend nodal noise to be phase-dependent seems unphysical. An exact description with Cauchy noise would therefore require a nontrivial extension of the framework and is left for future work. We remind however, that common noise can be included exactly, since the coefficients $(a,b,c)$ can be arbitrary time-dependent functions, allowing for example, the inclusion of shot noise to model finite-size effects~\cite{klinshov_kirillov_2022}.

With minor adjustments, ensembles with a finite number of two-phase QIF neurons can also be described in a low-dimensional way, closely analogous to the Watanabe–Strogatz (WS) theory~\cite{watanabe_strogatz_1993,watanabe_strogatz_1994}.
As in the WS framework, coupling introduces constants of motion into the macroscopic dynamics, which can complicate analytical treatment and slightly increase numerical cost.

The approach naturally generalizes to any number of voltage phases by partitioning the standard QIF trajectory into segments, each following its own Riccati dynamics.
In general, any M\"obius transformation of the voltage can define a new phase, though only specific forms --- such as Eq.~\eqref{eq:transform} --- ensure continuity at the phase boundaries.
Multiple phases could also be applied to complex Riccati equations~\cite{cestnik_martens_2024,pazo_cestnik_2025}, 
limiting the microscopic states to a region of the complex plane (e.g. inside/outside the unit disk).

\bibliography{bibliografy}

\end{document}